\documentclass[9pt,twocolumn,twoside]{opticajnl}
\journal{opticajournal} % use for journal or Optica Open submissions

\setboolean{shortarticle}{false}
\usepackage{lineno}
\usepackage{upgreek}
\title{Dual‑Wavelength Cancellation of Dispersion‑Induced Phase Noise in Opto‑Terahertz Fiber Links}
\author[1,*]{Brendan M. Heffernan}
\author[1]{James Greenberg}
\author[1]{William F. Mcgrew}
\author[1]{Antoine Rolland}

\affil[1]{Boulder Research Lab, IMRA America, Inc. 1551 S. Sunset St. Suite C, Longmont, CO 80501, USA}

\affil[*]{bheffern@imra.com} %% email address is required; see note below about the corresponding author designation

\begin{abstract} 
Stable dissemination of terahertz (THz) signals over long distances is important for next-generation synchronization networks, radio astronomy, and high-capacity wireless systems. Optical fiber provides a low-loss platform for coherent frequency transfer; however, when a THz carrier is encoded as the difference between two optical wavelengths, chromatic dispersion introduces differential phase noise that degrades spectral purity. Here, we demonstrate phase-coherent distribution of opto-THz carriers over 38~km of standard single-mode fiber using a dual-wavelength Brillouin laser (DWBL) combined with a dual-channel round-trip noise-cancellation architecture. By extracting the differential phase noise between the two optical lines via a dual-channel round-trip measurement, dispersion-mediated phase fluctuations are compensated, and the intrinsic stability of the source is effectively preserved at the remote end within the measurement sensitivity. Opto-THz carriers at 150, 300, and 600~GHz exhibit sub-femtosecond timing stability and fractional frequency instabilities below $10^{-17}$ at $10^{4}$~s of averaging.
\end{abstract}

\setboolean{displaycopyright}{false} % Do not include copyright or licensing information in submission.

\begin{document}

\maketitle

%%%%%%%%%%%%%%%%%%%%%%%%%%  body  %%%%%%%%%%%%%%%%%%%%%%%%%%
\section*{Introduction}

The dissemination of terahertz (THz) signals across long distances remains a central challenge for emerging applications in high-speed wireless communications \cite{Maekawa2024}, phase-coherent synchronization networks \cite{Meyer2025}, and large-scale scientific instrumentation such as X-ray free-electron lasers~\cite{Ulanov2025}. It is particularly relevant for radio astronomy  \cite{Shillue2011}, very-long-baseline interferometry (VLBI) \cite{Raymond2024}, and radio-over-fiber architectures, where carrier frequencies well above the microwave domain enable superior timing precision and enhanced data throughput. Conventional metallic waveguides, such as WR-3.4 for operation in the 220--330~GHz band, exhibit losses exceeding 14~dB/m and are prohibitively rigid and costly for deployment outside laboratory environments \cite{Gallot2000, Wang2004, Islam2020}. Free-space THz links can span hundreds of meters, but their practicality is limited by stringent alignment requirements, weather-dependent attenuation, and atmospheric absorption features that can severely constrain long-haul operation \cite{Schneider2012}.

A widely used approach is to encode the THz frequency as the difference between two optical wavelengths, forming an \emph{opto-THz} carrier, which can be recovered at the destination by optical heterodyning. Optical fiber provides a suitable distribution platform, offering sub-dB/km attenuation, electromagnetic isolation, and a mature global infrastructure developed for telecommunication networks \cite{Agrawal2022}. This approach leverages optical coherence while preserving the desired THz frequency in the electrical domain. Several groups have demonstrated this as a viable strategy over deployed fiber links extending to 90 km \cite{Kumagai2014} and laboratory-based recirculating loops of several thousand kilometers \cite{Sakaguchi2025}. However, the environment to which the fiber link was subjected strongly affected the fidelity of the distributed frequency \cite{Kumagai2014}. Achieving high precision therefore requires active noise cancellation.

Frequency dissemination over optical fiber is commonly implemented using a local station, a fiber link, and a remote station. The local station contains the source and noise-cancellation system, while the fiber link introduces phase noise due to environmental perturbations. To enable noise cancellation, a portion of the optical signal is returned from the remote station to the local station, allowing the accumulated phase noise to be measured and compensated. This round-trip stabilization technique has been widely used for the dissemination of ultra-stable optical frequencies \cite{Ma1994, Droste2013, Stefani2015}. In the context of terahertz generation, distribution of single optical tones followed by local synthesis has also been demonstrated, but typically requires multiple self-referenced frequency combs at the remote site \cite{Nagano2016}.

Feedback correction to a single optical frequency as part of opto-THz distribution has also been performed. The photonic local oscillator system of the Atacama Large Millimeter-submillimeter Array (ALMA) demonstrated distribution of opto-THz waves from 27 - 122 GHz over tens of kilometers using noise-canceled optical links~\cite{Shillue2011,Shillue2013}. However, stabilization of a single optical frequency does not suppress dispersion-induced differential phase noise when the THz carrier is generated from two widely spaced optical lines. Environmental perturbations modify the phases of the two wavelengths differently, leading to excess phase noise that degrades the spectral purity of the recovered THz signal.

In this work, we demonstrate opto-THz dissemination over 38~km of standard single-mode fiber using a dual-wavelength Brillouin laser (DWBL) combined with a dual-channel round-trip noise-cancellation scheme. Each optical line is independently frequency-shifted, enabling reconstruction of the phase fluctuations experienced by both wavelengths along the fiber. This architecture compensates both path-length (Doppler-like) noise and dispersion-induced group delay fluctuations in real time. Using out-of-loop electro-optic (EO) comb measurements at both ends of the link, we confirm suppression of fiber-induced phase noise and preservation of the source stability at the remote end. Opto-THz carriers at 150, 300, and 600~GHz exhibit sub-femtosecond timing stability and fractional frequency instabilities below $10^{-17}$ at $10^{4}$~s of averaging. This work complements previous demonstrations of opto-THz dissemination \cite{Sun2014, Wang2023, Li2023a} by providing complementary studies of the phase noise introduced by the fiber link, systematic frequency shifts cause by temperature change, and a discussion of the group index-mediated frequency fluctuations. 

\section*{Theoretical background}

When two optical wavelengths propagate through a common medium, environmental noise imprinted on the optical paths cancel to first order when the two fields are heterodyned. However, some noise remains. In practice, chromatic dispersion leads to differential phase accumulation.

In the dual-wavelength configuration considered here, the recovered phase of the opto-THz carrier depends on the \emph{difference} of the optical phase delays,
\begin{equation}
\phi_{\mathrm{THz}} =  L\bigl[\beta(\omega_{2})-\beta(\omega_{1})\bigr]\,
\end{equation}
where $\beta(\omega)$ is the propagation constant. Expanding $\beta$ about $\omega_1$ to first order in frequency separation, neglecting higher-order dispersion terms, yields
\begin{equation}
\phi_{\mathrm{THz}} \approx  L \frac{d\beta}{d\omega}(\omega_2 - \omega_1) = 2 \pi \frac{L}{v_{g}} \nu_{\mathrm{opto\text{-}THz}}
\label{eq:phi_thz}
\end{equation}
where $d\beta/d\omega = 1/v_g = n_g/c$, $v_g$ is the group velocity, and $n_g$ is the group index. \eqref{eq:phi_thz} suggests the opto-THz phase depends on the length of fiber, the opto-THz frequency, and the group velocity of the wavelengths through the medium. This is consistent with laser pulse propagation, where group index contributes to the overall delay time of a pulse traversing a medium and therefore of the timing of the electric field peak, otherwise known as timing jitter \cite{Agrawal2022}. In this case, timing jitter can be seen as phase noise of the resulting terahertz wave, because it results in time variation of the zero crossing of the wave. 

Examining contributions to fluctuations in the opto-THz phase $\delta\phi_{THz}$:
\begin{equation}
\delta \phi_{\mathrm{THz}}
= 2 \pi \frac{L}{v_g} \nu_{\mathrm{opto\text{-}THz}} \left[\frac{\delta L}{L}  - 
\frac{\delta v_g}{v_g}  +
\frac{\delta \nu_{\mathrm{opto\text{-}THz}}}{\nu_{\mathrm{opto\text{-}THz}}}
\right]
\label{eq:phi_thz_fluct}
\end{equation}
where the negative sign reflects the inverse dependence of phase on group velocity, as faster propagation reduces the accumulated delay. \eqref{eq:phi_thz_fluct} shows that phase fluctuations depend on variations in fiber length, group velocity, and carrier frequency. Environmental variations in temperature or strain modify the length and group index of the fiber link. In the case of temperature fluctuations, we can distinguish the relative strength of each term by considering its fractional change. Silica expands by approximately $4\times 10^{-7} \mathrm{K}^{-1}$ \cite{Souder1925} while group velocity changes by $\sim 8\times 10^{-6} \mathrm{K}^{-1}$ (based primarily on the thermo-optic coefficient of silica) \cite{Leviton2006, Rego2023}. This analysis indicates that group velocity fluctuations are expected to dominate the excess phase noise under typical thermal perturbations. 

\section*{Experimental setup}
\begin{figure*}[tb]
\includegraphics[width=7in]{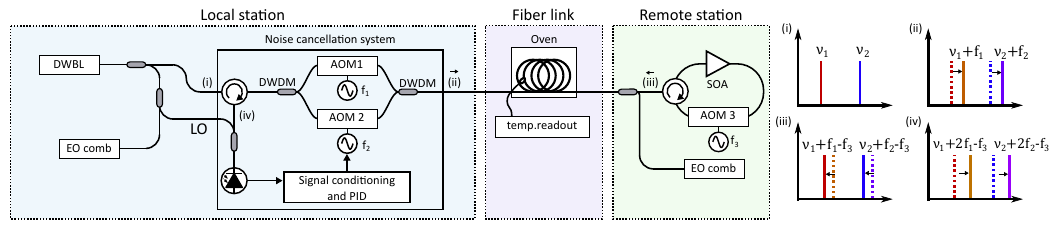}
\caption{\label{figure1}Schematic overview of the opto-THz frequency distribution test setup and associated noise cancellation system. Depictions of the optical frequencies at several points are denoted using Roman numerals, showing the result of various frequency shifts. Arrows above the Roman numerals denote the direction the light is traveling. DWBL; dual-wavelength Brillouin laser, EO comb; electro-optic comb, LO; local oscillator, DWDM; dense wavelength division multiplexer, AOM; acousto-optic modulator, PID; proportional, integral, derivative control system, SOA; semiconductor optical amplifier.}
\end{figure*}

As mentioned previously, the distribution and noise-cancellation architecture operates by measuring the frequency noise added by the fiber link and feeding back the opposite noise to remove it at the remote station. A schematic overview of how this is achieved is shown in Fig.~\ref{figure1}. As a broad overview, the local station consisted of a two-wavelength laser, acousto-optic modulators (AOMs) for frequency control, and electronics for noise-cancellation. The fiber link consisted of 38~km of standard single-mode fiber (SMF-28) spooled inside a laboratory oven (Thermo Scientific Heratherm). The oven provided a moderate amount of passive thermal isolation from room fluctuations but remained unpowered unless stated otherwise. The remote station provided optical amplification, frequency tagging, and a return path to enable round-trip sensing. 

More specifically, a dual-wavelength Brillouin laser (DWBL)~\cite{Heffernan2024} served as the optical/opto-THz source. Its exceptionally low phase noise and long coherence length enable optical self-heterodyne comparisons after propagation through 38~km (76~km round-trip) of fiber. Based on the fractional frequency instability reported in~\cite{Heffernan2024}, the opto-THz carrier at 300~GHz exhibits an integrated linewidth of less than 10~Hz~\cite{Domenico2010}, corresponding to a THz-band coherence length exceeding 10,000~km. Using the measured optical phase-noise spectra \cite{Egbert2025}, the individual optical lines are estimated to possess coherence lengths greater than 800 km. These coherence properties ensure that the returned signal remains phase-coherent with the local reference over the full round-trip delay of 368~$\upmu$s. In addition to possessing high coherence, the DWBL is also widely tunable, enabling characterization at 150, 300, and 600 GHz opto-THz frequencies. 

Light from the DWBL was split, with 99\% routed to the frequency-control elements and subsequently injected into the fiber link. The remaining 1\% provided the local oscillator (LO) and was split 50:50. Half was used for heterodyne readout to extract link noise and half was used for EO comb frequency readout~\cite{Rolland2011}. The two DWBL wavelengths (frequencies $\nu_{1}$ and $\nu_{2}$) were separated using dense-wavelength-division multiplexers (DWDMs) and independently frequency-shifted by acousto-optic modulators (AOM1 and AOM2). AOM1, driven at $f_{1}$ by a frequency synthesizer, produced an output at $\nu_{1}+f_{1}$. AOM2, driven by a voltage-controlled oscillator (VCO) at $f_{2}$, produced light at $\nu_{2}+f_{2}$. After recombination, the resulting opto-THz frequency was
\[
\nu_{\mathrm{opto\text{-}THz}} = (\nu_{2}+f_{2}) - (\nu_{1}+f_{1}).
\]

\noindent This opto-THz wave was then sent through the fiber link to the remote station.

At the remote station, the incoming light was split, with a fraction used for EO comb frequency readout. The remaining portion was sent through a three port circulator, amplified using a semiconductor optical amplifier (SOA), shifted by AOM3 (driven at frequency $f_{3}$), and then returned to the fiber link. The $-1$ diffracted order was selected so that both optical lines incurred a common shift of $-f_{3}$. This tagging step does not change the opto-THz frequency, as both optical lines experience the same frequency shift but uniquely identifies light that has traversed the entire link, distinguishing it from backscattered (Rayleigh-reflected) components \cite{Agrawal2022}.

The frequency-tagged light returned to the local station and passed through the same DWDM/AOM chain a second time. After exiting the circulator, it was combined with the LO and detected on a 150-MHz bandwidth photodiode (PD). The optical tones incident on the PD were
\[
\{\nu_{1},\, \nu_{2},\, \nu_{1}+2f_{1}-f_{3},\, \nu_{2}+2f_{2}-f_{3}\},
\]
giving rise to RF mixing products at $2f_{1}-f_{3}$ and $2f_{2}-f_{3}$. These RF tones carried the sum of the forward and backward link noise for each optical wavelength. Because the round-trip delay is 368~$\upmu$s, link noise below 2.7 kHz is largely common to both directions and can therefore be effectively corrected by the servo. The frequencies of the two mixing products can therefore be written as 
\[
2f_{1}-f_{3} + 2N_{\mathrm{fiber}, \nu_1} \;\;\mathrm{and}\;\; 2f_{2}-f_{3} + 2N_{\mathrm{fiber}, \nu_2}
\]
where $N_{\mathrm{fiber}, \nu_j}$ represents the mean-zero frequency fluctuations caused by fiber noise at Fourier frequencies below 2.7 kHz, as they affect optical frequency $\nu_j$. A narrow bandpass filter isolated the two mixing products, which were then jointly amplified. The amplifier nonlinearity was intentionally exploited to produce a difference frequency containing $2(f_{2}-f_{1})$ plus twice the differential link noise $2(N_{\mathrm{fiber},\nu_2}-N_{\mathrm{fiber},\nu_1})$, where the factor of two arises from the round-trip accumulation of phase noise. After filtering and dividing by two, the resulting IF signal was
\[
f_{\mathrm{IF}} = (f_{2}-f_{1}) + (N_{\mathrm{fiber},\nu_2} - N_{\mathrm{fiber},\nu_1}),
\]
which encodes the differential phase noise of the fiber link affecting the opto-THz carrier (see Appendix A for a detailed derivation).

An FPGA-based instrument (Moku:Go) served as the synthesizer, mixer, and PID controller. Its error signal was used to adjust the VCO driving AOM2, thereby compensating the measured link-induced phase fluctuations in real time (see Appendix A for the closed loop formulation). Typical values in the experiment were $f_{1}=83.5$~MHz, $f_{2}=78.2$~MHz, and $f_{3}=74.8$~MHz, with optical frequencies near $\nu_{2}\approx 193.4$~THz and opto-THz spacings of 150, 300, and 600~GHz. All synthesizers and the VCO were referenced to a common frequency standard.

\section*{Results}
\subsection*{Free-running fiber-link instability}
Electro-optic (EO) comb readouts~\cite{Rolland2011} at both ends of the link measured the opto-THz frequency before and after transmission. The frequencies at both stations recorded simultaneously using a dead-time-free frequency counter referenced to a rubidium frequency standard. These EO combs served exclusively as diagnostic instruments and were not involved in the stabilization process. 

\begin{figure}[ht!]
\centering\includegraphics[width=3.4in]{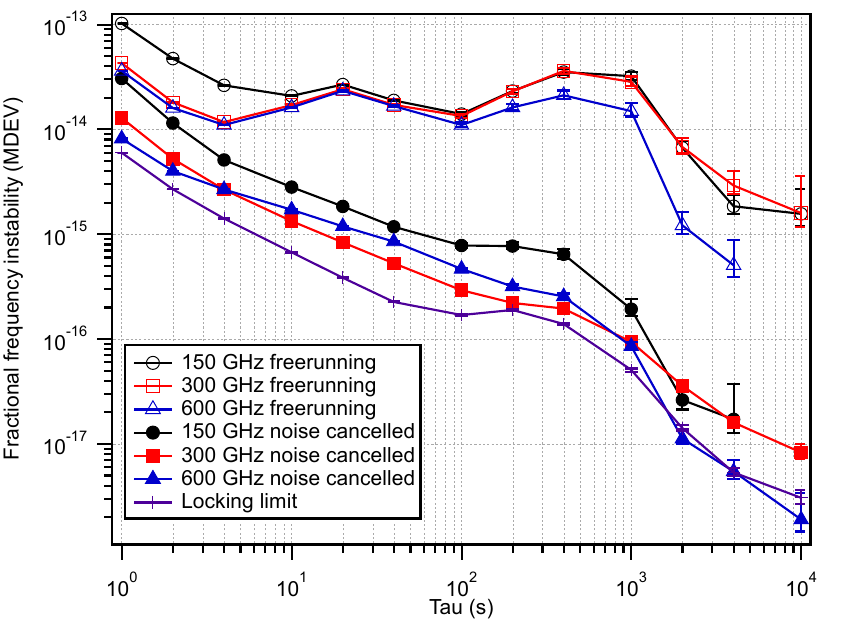}
\caption{\label{fig:mdevs}Fractional frequency instability, in terms of modified Allan deviation (MDEV) of the 38~km fiber link in free-running (open symbols) and noise-canceled (filled symbols) operation at 150~GHz, 300~GHz, and 600~GHz. The free-running curves display a flicker-frequency floor at the low-to-mid $10^{-14}$ level, whereas stabilization reduces the instability below $10^{-17}$ at long averaging times. The MDEV of the in-loop noise, measured at $\nu_{opto-THz} = 600 \mathrm{GHz}$ is plotted in purple, showing the locking limit of our experiment.}
\end{figure}

Before activating the stabilization system, we characterized the intrinsic frequency noise introduced by the 38~km fiber link under conditions where both optical wavelengths co-propagate along a common path. The noise-cancellation loop was bypassed, and the 99\% output of the source splitter was launched directly into the fiber. This configuration avoids differential path-length fluctuations between the two wavelengths prior to the link, which would otherwise convert uncorrelated optical phase noise into opto-THz phase noise. Because the heterodyne process maps differential optical phase fluctuations directly onto the THz carrier without frequency division, even small uncorrelated optical phase perturbations lead to disproportionately large fractional frequency noise at THz frequencies. Figure~\ref{fig:mdevs} shows the resulting fractional frequency instability, evaluated using the modified Allan deviation (MDEV), for opto-THz carrier frequencies of 150~GHz, 300~GHz, and 600~GHz.

In all three cases, the link exhibits a characteristic flicker-frequency floor at the low-to-mid $10^{-14}$ level for averaging times from $10^{0}$~s to $10^{2}$~s. This behavior is consistent with thermally driven group index fluctuations in standard SMF, which dominate fiber-link instability in the absence of active cancellation.

\subsection*{Stabilized link performance}

When the dual-wavelength round-trip noise-cancellation system is activated, the instability improves by more than three orders of magnitude. As shown in Fig.~\ref{fig:mdevs}, the residual instability after cancellation decreases as $\tau^{-1}$ for averaging times up to approximately $10^{4}$~s, reaching fractional frequency instabilities below $10^{-17}$ for all three opto-THz carriers. This behavior is characteristic of white phase noise dominated by the residual error of the servo and the noise floor of the optical source itself.

No flicker floor is observed down to $10^{-17}$, demonstrating that the stabilization successfully suppresses both path-length noise and group velocity fluctuations. The agreement among the three THz carriers further confirms that the servo corrects the fundamental differential group velocity responsible for the free-running degradation.

\subsection*{Effect of link temperature change on frequency}
We evaluated the ability of our noise cancellation system to hold through systematic temperature changes applied to the fiber link by comparing the opto-THz frequencies at the local and remote stations. We began the measurements by monitoring for 10 minutes with the link oven at a baseline temperature. We then increased the setpoint of the oven temperature and waited 25 minutes for the temperature to ramp and equilibrate. The oven door was then opened to rapidly cool the oven to room temperature. A thermistor placed between winds of fiber was used as an independent temperature readout throughout the experiment. 

The response of the free-running link is shown in Fig.~\ref{fig:temp_ramp}. The measured temperature (red, right axis) is plotted alongside the frequency difference between the local and remote stations (black, left axis). The observed frequency shift does not follow the absolute temperature, but rather its temporal variation. This behavior arises because temperature-induced changes in optical path length and group delay contribute to the accumulated phase, while the measured frequency corresponds to its time derivative,
\begin{equation}
\delta \nu(t) = \frac{1}{2\pi}\frac{d}{dt}\,\delta \phi(t).
\end{equation}
As a result, temperature variations produce frequency shifts proportional to $dT/dt$.

A temperature change of approximately $2^{\circ}\mathrm{C}$ over $\sim 10^{3}$~s induces a frequency shift of $\sim 1$~Hz at a 600~GHz opto-THz carrier, corresponding to a fractional frequency shift of $1.7 \times 10^{-12}$. This exceeds the flicker floor of $\sim 3 \times 10^{-14}$ shown in Fig.~\ref{fig:mdevs}. The measurement was repeated three times to confirm reproducibility (not shown), and similar measurements were performed at 150~GHz and 300~GHz. The frequency shift scales linearly with carrier frequency, indicating that the perturbation is phase-based and leading to a consistent fractional frequency shift.

We performed a similar routine while the fiber-noise-cancellation loop was active. The temperature (Fig. \ref{fig:temp_ramp}, red) ramps in a similar manner but is left longer to equilibrate before opening the oven door to perform a quick temperature shock. However, the feedback loop is able to stabilize the frequency difference (Fig. \ref{fig:temp_ramp}, black line) such that minimal deviation is observed. The loop holds even for fast temperature changes of $4^{\circ}\mathrm{C}$.  

\begin{figure}[ht!]
\centering\includegraphics[width=3.4in]{fig_temp_pulse_alt1.pdf}
\caption{\label{fig:temp_ramp}Frequency difference between the opto-THz reference at the local and remote stations while link temperature was changed. Top: free-running operation. Bottom: while the noise cancellation system was active. }
\end{figure}

\subsection*{Phase noise added by fiber link}
\begin{figure}[ht!]
\centering\includegraphics[width=3.4in]{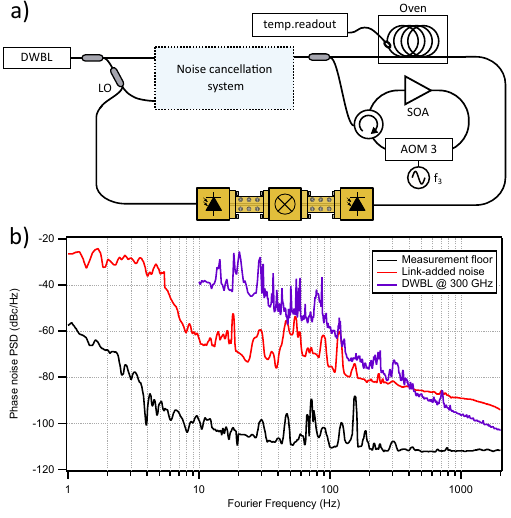}
\caption{\label{fibernoise}a) System used to measure the additive phase noise of the fiber link. b) Resulting single-sideband phase noise spectra. The residual phase noise of the fiber link (red) is well above the noise floor (black), as measured by a 300 GHz opto-THz carrier. Delay-line effects limit the results to below ~2 kHz. The phase noise of the DWBL reference (described in \cite{Egbert2025}) is plotted in purple.}
\end{figure}

The phase noise added by the fiber link was studied using the setup shown in Fig. \ref{fibernoise} a). To measure the residual noise of the link, we radiated the DWBL LO using a  uni-travelling carrier photodiode (UTC-PD) \cite{Ishibashi1997, Ishibashi2020}. This signal was mixed on a fundamental mixer against the opto-THz wave that traversed the fiber link, producing a beat signal at $((\nu_1+f_1)-(\nu_2+f_2)) - (\nu_1 -\nu_2) = f_1-f_2$. This RF signal was then amplified and analyzed by a phase noise analyzer (Microchip 53100A).

The noise cancellation system shown in Fig. \ref{fibernoise} a) is identical to that shown in Fig. \ref{figure1}. It is needed in order to generate a beat note for phase noise analysis. However, splitting the two wavelengths of the DWBL and modulating them separately results in non-common noise which does not cancel at photodetection. To ensure that the residual phase noise measured is only that of the fiber link, we split the light coming from the noise cancellation system and used half to send back through the AOMs of the cancellation system and engaged the system electronics. In this way, we effectively canceled noise associated with the noise cancellation system itself and are able to quantify the residual noise of the fiber link in isolation.

To quantify the level of noise cancellation achieved, we measured the LO against the opto-THz coming directly before the fiber link (or equivalently bypassed the 38 km of fiber shown in Fig. \ref{fibernoise} a).) Results are shown in Fig. \ref{fibernoise} b). The level of noise cancellation prior to sending the signal through the fiber is shown in black. It is effectively the measurement floor of the residual phase noise measurement. The residual phase noise of the 38 km fiber link under our laboratory conditions is plotted in red. Fourier frequencies above $\approx2$ kHz are affected by the transfer function of the fiber delay (fiber link) and are therefore unreliable \cite{Kuse2017}. To put these results in context, we plot a representative phase noise trace for the DWBL used in this study in purple (reproduced from \cite{Egbert2025}). 

The phase noise added by the fiber link above 10 Hz has a characteristic $1/f$ slope indicating phase flicker. This background slope is in addition to spurs which are likely a result of vibrational noise. In our case, we expect that deleterious phase noise will be added above $\approx$400 Hz. It is evident from this measurement that phase noise from the contributions outlined in Eq. \ref{eq:phi_thz_fluct} will become non-negligible at Fourier frequencies near the bandwidth limitations of the loop, set by the physical link length. Therefore, we conclude that slower variations caused by environmental variables are the most relevant for frequency dissemination and that fiber-induced noise at higher Fourier frequencies appears to be comparatively small relative to the source noise under our experimental conditions.

\section*{Discussion}
The dual-wavelength, round-trip cancellation scheme robustly suppresses the environmental noise of a 38~km fiber link. The resulting remote opto-THz carriers at 150, 300, and 600~GHz maintain the intrinsic instability of the DWBL, reaching values below $10^{-17}$ at $10^{4}$~s of averaging time. These results compare favorably with other reports using similar noise-canceling schemes \cite{Sun2014, Wang2023, Li2023a}. 

We have found theoretically that noise sources in the fiber, primarily caused by fluctuations in the environment, are mediated through the group velocity. This matches well with intuition from the field of ultra-fast optics, where group velocity/index is directly tied to pulse arrival time and timing jitter. Furthermore, we estimate that changes in the group index are likely the dominant contribution to frequency and phase noise under the conditions considered.

The phase noise added by a fiber link onto an opto-THz wave was measured by canceling noise at the point directly before the fiber link, removing the effects of non-common paths required for source preparation. We have found a 1/f functionality at Fourier frequencies greater than 10 Hz. Assuming this functionality holds, it suggests that the 38 km of fiber contributes approximately 2.9 femtoseconds of timing noise jitter when integrated from 10 Hz to 10 MHz (.25 as/GHz/km). This analysis neglects spurs caused by vibrations as they may vary from link to link. If these effects are included, the link shown here contributes 9.6 femtoseconds of timing noise jitter (.8 as/GHz/km).

Note that polarization has been neglected in this study. All components used were polarization maintaining except for the fiber link. In principle, this can cause signal deterioration due to polarization rotation in the link, particularly over many hours or days. Mitigation strategies such as a polarization randomizer on the optical source\cite{Wang2023} or automated polarization control might be employed, but were not necessary for, and not the focus of, the work presented here.

The present implementation relies on a single fiber span under laboratory conditions. In deployed networks, temperature excursions and strain fluctuations may be larger, and the chromatic-dispersion coefficient can vary with environmental conditions and aging. Nonetheless, the architecture demonstrated here is directly compatible with installed telecommunication fibers and can be extended to include bidirectional amplifiers, wavelength-division multiplexing, or active environmental compensation. Because the technique operates entirely in the optical domain, it can be adapted to other dual-wavelength sources, including dual-mode microcombs, optical parametric oscillators, or integrated semiconductor lasers.

\appendix
\section{Appendix: Differential phase-noise extraction and closed-loop suppression}

In this Appendix, we derive how the round-trip measurement extracts the differential phase fluctuations between two optical wavelengths and how feedback applied through AOM2 suppresses the resulting opto-THz phase noise. The derivation applies to all differential phase noise accumulated along the optical paths, including contributions from the fiber link, discrete optical components, and any non-common optical paths. The analysis is restricted to Fourier frequencies below the inverse round-trip delay, such that forward- and backward-propagating perturbations may be treated as approximately equal.

\subsection*{Optical fields and single-pass phase noise}

Let the two optical fields launched into the system be
\begin{equation}
E_1(t) = A_1 e^{i 2\pi (\nu_1 + f_1)t}, 
\qquad
E_2(t) = A_2 e^{i 2\pi (\nu_2 + f_2)t},
\end{equation}
where $\nu_1$ and $\nu_2$ are the two optical frequencies and $f_1$ and $f_2$ are the frequency shifts applied at the local station.

After propagation through the optical system, each field acquires a phase perturbation
\begin{equation}
\phi_{j}(t), \qquad j \in \{1,2\},
\end{equation}
which includes all accumulated phase noise along its optical path.

At the remote station, the heterodyne phase of the opto-THz carrier is
\begin{equation}
\phi_{\mathrm{THz}}(t)
=
2\pi \big[(\nu_2+f_2)-(\nu_1+f_1)\big]t
+
\phi_{2}(t)-\phi_{1}(t).
\label{eq:appendix_phi_thz_remote_rewrite}
\end{equation}
The noise contribution is therefore the \emph{differential phase}
\begin{equation}
\phi_{\mathrm{diff}}(t)
\equiv
\phi_{2}(t)-\phi_{1}(t),
\label{eq:appendix_phi_diff_def_rewrite}
\end{equation}
which includes both common-medium effects (e.g., fiber dispersion) and non-common-path contributions (e.g., wavelength-dependent routing through optical components).

\subsection*{Round-trip phase measurement}

At the remote station, both optical fields are shifted by $-f_3$ and sent back through the same optical paths. After a second pass through the local AOM chain, the returned fields generate RF tones at
\begin{equation}
2f_1-f_3,
\qquad
2f_2-f_3,
\end{equation}
when beaten against the local oscillator fields.

Including accumulated phase noise, these signals can be written as
\begin{align}
V_1(t) &\propto \cos\!\left[2\pi(2f_1-f_3)t + \phi^{\mathrm{rt}}_{1}(t)\right],\\
V_2(t) &\propto \cos\!\left[2\pi(2f_2-f_3)t + \phi^{\mathrm{rt}}_{2}(t)\right],
\end{align}
where $\phi^{\mathrm{rt}}_{j}(t)$ denotes the round-trip phase fluctuation for each wavelength.

For Fourier frequencies below the inverse round-trip delay, the forward and backward perturbations are approximately equal, giving
\begin{equation}
\phi^{\mathrm{rt}}_{j}(t) \approx 2\,\phi_{j}(t).
\end{equation}
Taking the difference yields
\begin{align}
\phi_{\mathrm{err}}(t)
&=
\phi^{\mathrm{rt}}_{2}(t)-\phi^{\mathrm{rt}}_{1}(t)
\nonumber\\
&\approx
2\big[\phi_{2}(t)-\phi_{1}(t)\big]
\nonumber\\
&=
2\,\phi_{\mathrm{diff}}(t).
\end{align}
After appropriate scaling, the measured error signal is therefore
\begin{equation}
\phi_{\mathrm{meas}}(t) \approx \phi_{\mathrm{diff}}(t).
\end{equation}

This result is general: the round-trip measurement extracts the total differential phase noise accumulated between the two optical paths, independent of its physical origin.

\subsection*{Feedback through AOM2}

The feedback loop adjusts the phase of AOM2 to suppress the measured differential phase noise. Let $\phi_c(t)$ denote the applied correction.

The remotely recovered opto-THz phase becomes
\begin{equation}
\phi_{\mathrm{THz,res}}(t)
=
\phi_{\mathrm{diff}}(t) + \phi_c(t),
\end{equation}
while the round-trip error signal contains the correction twice,
\begin{equation}
\phi_{\mathrm{err}}(t)
\approx
2\big[\phi_{\mathrm{diff}}(t)+\phi_c(t)\big].
\end{equation}
The servo input is therefore
\begin{equation}
\phi_{\mathrm{meas}}(t)
\approx
\phi_{\mathrm{diff}}(t)+\phi_c(t).
\end{equation}

\subsection*{Closed-loop suppression}

In the Fourier domain, defining $\Phi_{\mathrm{diff}}(f)$, $\Phi_c(f)$, and $\Phi_{\mathrm{res}}(f)$, the closed-loop response is
\begin{equation}
\Phi_{\mathrm{res}}(f)
=
\frac{\Phi_{\mathrm{diff}}(f)}{1+L(f)},
\end{equation}
where $L(f)=G(f)H(f)$ is the open-loop transfer function.

Thus, all differential phase noise within the loop is suppressed by the sensitivity function
\begin{equation}
S(f)=\frac{1}{1+L(f)}.
\end{equation}

\subsection*{Effect of round-trip delay}

The round-trip delay $\tau_{\mathrm{rt}}$ introduces a phase factor $e^{-i2\pi f\tau_{\mathrm{rt}}}$ in the loop transfer function, limiting the effective bandwidth to
\begin{equation}
f_{\mathrm{BW}} \sim \frac{1}{\tau_{\mathrm{rt}}}.
\end{equation}
Below this frequency, the round-trip signal provides an accurate estimate of the single-pass differential phase noise. Above this frequency, forward and backward perturbations decorrelate, reducing the effectiveness of the correction.

\subsection*{Connection to group-delay fluctuations}

From the main text, the opto-THz phase may be written as
\begin{equation}
\phi_{\mathrm{THz}} \approx 2\pi \frac{L}{v_g}\nu_{\mathrm{opto\text{-}THz}}.
\end{equation}
Fluctuations in path length and group velocity produce
\begin{equation}
\delta \phi_{\mathrm{THz}}
=
2\pi \frac{L}{v_g}\nu_{\mathrm{opto\text{-}THz}}
\left(
\frac{\delta L}{L}
-
\frac{\delta v_g}{v_g}
\right).
\end{equation}
These contributions are encompassed within $\Phi_{\mathrm{diff}}(f)$ and are therefore suppressed by the feedback loop within its bandwidth.

\begin{backmatter}

\subsection*{Disclosures.} All authors: IMRA America, Inc. (E,P)

\subsection*{Data availability.} Data underlying the results presented in this paper are not publicly available at this time but may be obtained from the authors upon reasonable request.
%%%%%%%%%%%%%%%%%%%%%%% References %%%%%%%%%%%%%%%%%%%%%%%%%

%%%%%%%%%% If using BibTeX:
\bibliography{zotero_refs_all.bib}
\end{backmatter}
\end{document}